\documentclass[a4paper,11pt]{article}
\usepackage{jcappub}
\usepackage{amsmath}

\makeatletter
\gdef\@fpheader{}
\makeatother

\newcommand{\sss}[1]{{\scriptscriptstyle{#1}}}

\newcommand{\dd}{\mathrm{d}}
\newcommand{\ud}{\dd}
\newcommand{\ee}{\mathrm{e}}
\newcommand{\ue}{\ee}

\newcommand{\urad}{\mathrm{rad}}
\newcommand{\ueq}{\mathrm{eq}}
\newcommand{\ureh}{\mathrm{reh}}
\newcommand{\uend}{\mathrm{end}}

\newcommand{\unuc}{\mathrm{nuc}}

\newcommand{\uN}{\mathrm{N}}

\newcommand{\ux}{\mathrm{x}}

\newcommand{\calL}{\mathcal{L}}

\newcommand{\MeV}{\mbox{MeV}}
\newcommand{\GeV}{\mbox{GeV}}

\newcommand{\G}{\mathrm{G}}
\newcommand{\muG}{\mu\mathrm{G}}
\newcommand{\Mpl}{M_{_{\mathrm Pl}}}
\newcommand{\GN}{\mathrm{G_\uN}}

\newcommand{\Rrad}{R_\urad}

\newcommand{\Rx}{R_\ux}

\newcommand{\rhoreh}{\rho_\ureh}
\newcommand{\rhoend}{\rho_\uend}
\newcommand{\rhogamma}{\rho_\gamma}
\newcommand{\rhonuc}{\rho_\unuc}
\newcommand{\rhoB}{\rho_{\sss{B}}}
\newcommand{\rhoBzero}{\rho_{\sss{B_0}}}
\newcommand{\rhoBend}{\rho_{\sss{B_\uend}}}
\newcommand{\areh}{a_\ureh}
\newcommand{\zreh}{z_\ureh}
\newcommand{\zeq}{z_\ueq}
\newcommand{\azero}{a_0}
\newcommand{\aeq}{a_\ueq}
\newcommand{\aend}{a_\uend}
\newcommand{\zend}{z_\uend}
\newcommand{\Bzero}{B_0}
\newcommand{\Bend}{B_\uend}
\newcommand{\Nend}{N_\uend}
\newcommand{\Nreh}{N_\ureh}

\newcommand{\wreh}{\overline{w}_\ureh}

\newcommand{\OmegaR}{\Omega_\urad}

\title{Reheating constraints in inflationary magnetogenesis}

\author[]{Vittoria Demozzi and}
\author[]{Christophe Ringeval}

\affiliation[]{Centre for Cosmology, Particle Physics and Phenomenology,
  Institute of Mathematics and Physics, Louvain University, 2 Chemin
  du Cyclotron, 1348 Louvain-la-Neuve (Belgium)}

\emailAdd{vittoria.demozzi@uclouvain.be}
\emailAdd{christophe.ringeval@uclouvain.be}

\abstract{Among primordial magnetogenesis models, inflation is a prime
  candidate to explain the current existence of cosmological magnetic
  fields. Assuming conformal invariance to be restored after
  inflation, their energy density decreases as radiation during the
  decelerating eras of the universe, and in particular during
  reheating. Without making any assumptions on inflation, on the
  magnetogenesis mechanism and on how the reheating proceeded, we show
  that requiring large scale magnetic fields to remain subdominant
  after inflation gives non-trivial constraints on both the reheating
  equation of state parameter and the reheating energy scale. In terms
  of the so-called reheating parameter, we find that $\ln\Rrad >
  -10.1$ for large scale magnetic fields of the order $5 \times
  10^{-15}$ Gauss today. This bound is then compared to those already
  derived from Cosmic Microwave Background (CMB) data by assuming a
  specific inflationary model. Avoiding magnetic field backreaction is
  always complementary to CMB and can give more stringent limits on
  reheating for all high energy models of inflation. For instance, a
  large field matter dominated reheating cannot take place at an
  energy scale lower than typically $500\,\GeV$ if the magnetic field
  strength today is $\Bzero = 5 \times 10^{-15}\,\G$, this scale going
  up to $10^{10}\,\GeV$ if $\Bzero = 10^{-9}\,\G$.}


\begin{document}

\maketitle

\section{Introduction}
\label{sec:intro}

The origin of large scale magnetic fields in the universe is still a
mystery~\cite{Grasso:2000wj}. There are evidences of their presence in
the intergalactic medium and this strongly suggests that the origin of
these fields might be primordial~\cite{Giovannini:2003yn,
  Kandus:2010nw}. Magnetic fields in galaxies have a strength ranging
from $5$ to $100\,\muG$~\cite{Chyzy:2004hr} while the strength
detected within clusters of galaxies is of the order $1$ to
$10\,\muG$~\cite{Vogt:2003su}. Moreover limits on magnetic fields in
the intergalactic medium have been recently derived using combined
data from the Atmospheric Cherenkov Telescopes and the Fermi Gamma-Ray
Space Telescope on the spectra of distant blazars. Assuming blazars
are injecting both gamma and cosmic rays, Ref.~\cite{Essey:2010nd}
reports the two-sigma confidence interval from $1 \times 10^{-17}\,
\G$ to $3 \times 10^{-14} \, \G$. Without assuming cosmic rays
production, solely the lower limit remains model
independent~\cite{Essey:2010nd}. Other data from HESS and Fermi
imposes a lower bound of $5\times 10^{-15}\, \G$~\cite{Dolag:2010ni,
  Neronov:1900zz, Tavecchio:2010mk} while Faraday rotations give an
upper limit as $10^{-9}\,\G$~\cite{Kronberg:1993vk}.

Inflation is a candidate of choice to generate primordial magnetic
fields~\cite{Turner:1987bw}, because, as for large scale structures
and Cosmic Microwave Background (CMB) anisotropies, it provides the
means of producing effects on scales larger than Hubble radius today,
starting from processes on much smaller length scales. As for
curvature perturbations, the idea is the amplification of quantum
vacuum fluctuations of the electromagnetic field during
inflation. However this is not straightforward since the standard
electromagnetic Lagrangian $\calL = - \frac{1}{4} F^{\mu\nu} F_{\mu
  \nu}$ is conformally invariant. As a result, for a comoving observer
$u_\mu$ in a flat Friedmann-Lema\^{\i}tre universe, magnetic fields
$B^\mu \equiv -\frac{1}{2} \varepsilon^{\mu \nu \alpha \beta}
F_{\alpha \beta} \, u_\nu$ always decrease as $1/a^{2}$, where $a$ is
the scale factor. Notice that such an evolution is modified in an open
universe~\cite{Barrow:2011ic}. In a flat universe, this problem can be
solved by breaking the conformal invariance of the electromagnetic
field during inflation. Several mechanisms have been proposed to do
so~\cite{Turner:1987bw, Ratra:1991bn, Dolgov:1993mu, Dolgov:1993vg,
  Ratra:1994vw, Tornkvist:2000js, Gasperini:1995dh, Lemoine:1995vj,
  Ashoorioon:2004rs, Bamba:2006ga, Martin:2007ue, Durrer:2010mq,
  Barnaby:2012tk}, some of them producing observable non-Gaussianities
in the CMB~\cite{Barnaby:2012tk, Bonvin:2011dt}.
  
Once conformal invariance is broken, and the field is amplified,
another issue that one should consider is the backreaction of the
generated field onto the background. Already during inflation,
backreaction can be problematic as the electromagnetic stress can
overcome the inflaton energy density and this gives very strong bounds
on the maximum strength of the primordial fields~\cite{Demozzi:2009fu,
  Kanno:2009ei, Byrnes:2011aa, Finelli:2011cw, Urban:2011bu,
  Koivisto:2011rm}. Moreover, the time evolution of the
electromagnetic stress during inflation depends on the way conformal
invariance is broken thereby rendering the backreaction problem model
dependent.

Nevertheless, after the end of inflation, mostly all primordial
magnetogenesis models assume that conformal invariance is restored
such that, without source, the magnetic field decreases inversely
proportional to the scale factor squared. This will be our working
hypothesis in the following. Let us however mention that magnetic
fields could also be affected by other mechanisms occurring after
inflation~\cite{DiazGil:2007dy, Jedamzik:2010dq, Easther:2010mr}, but
these are expected to affect length scales not much larger than the
Hubble radius at that time. As a result, focusing on super-Hubble
modes, their associated energy density $\rhoB=B^2/2$ still behaves as
radiation and the strength of the magnetic field today is simply
redshifted since the end of inflation
\begin{equation}
\label{eq:Bzero}
\Bzero = \dfrac{\Bend}{(1 + \zend)^2}\,.
\end{equation}
Here $\Bzero$ and $\Bend$ are respectively the magnetic field on
Hubble scale today and at the end of inflation while $\zend
=\azero/\aend-1$ is the redshift at which inflation ended. In this
paper, we point out that $\zend$ depends on the properties of
reheating~\cite{Turner:1983he, Kofman:1997yn, Garcia-Bellido:1997wm,
  Liddle:2003as, Desroche:2005yt} and therefore the value of magnetic
fields today is connected to the reheating epoch for all models of
inflationary magnetogenesis. In particular we show that requiring
large scale magnetic fields not to have backreaction \emph{after} the
end of inflation, yields a lower bound on the reheating parameter $\ln
\Rrad > -10.1$ for $B_0=5\times 10^{-15}\,\G$. The reheating parameter
has been introduced in Refs.~\cite{Martin:2006rs, Ringeval:2007am,
  Martin:2010kz} and can be expressed under various equivalent forms
such as
\begin{equation}
\label{eq:lnRrad}
\ln \Rrad = \frac{\Delta N}{4}\left(-1+3 \wreh \right) = \frac{1-3
  \wreh}{12 (1+\wreh) } \ln \left(\frac{\rhoreh}{\rhoend} \right).
\end{equation}
Here ``reh'' and ``end'' stand respectively for the end of reheating
and inflation while $\wreh$ is the \emph{mean} equation of state
parameter during reheating. The quantity $\Delta N
=\Nreh-\Nend=\ln(\areh/\aend)$ is the number of e-folds reheating
lasted. Requiring to be consistent with standard cosmology, namely
that reheating occurs before Big-Bang Nucleosynthesis (BBN) and after
inflation, one gets $\rhonuc\equiv(10\,\MeV)^4<\rhoreh<\rhoend <
(10^{-5}\Mpl)^4$. The upper limit comes from the observed amplitude of
the CMB anisotropies\footnote{$\Mpl\equiv 1/\sqrt{8\pi\GN}$ is the
  reduced Planck mass.}. Moreover, the positivity energy conditions in
General Relativity impose that $-1/3<\wreh<1$ such that the reheating
parameter could take any value in the range $\ln \Rrad \in [-35,12]$.
Our result, $\ln \Rrad > -10.1$, is therefore a non-trivial lower
bound. As Eq.~(\ref{eq:lnRrad}) emphasizes, this limit can be
propagated either into a lower bound on the energy density of
reheating, or number of e-folds, if one assumes the equation of state
to be known; or the converse.

Let us mention that the seven year Wilkinson Microwave Anisotropies
Probe (WMAP7) CMB data have been shown to constrain the reheating
parameter, but provided an inflationary model is
specified~\cite{Komatsu:2010fb, Larson:2010gs, Jarosik:2010iu,
  Martin:2010kz, Martin:2010hh, Mortonson:2010er, Easther:2011yq}. Our
result does not need this assumption and will be compared to CMB in
Sec.~\ref{sec:cmbreh}.

Equation~(\ref{eq:Bzero}) is valid at any length scale not affected by
post-inflationary mechanisms and more recent astrophysical
processes. Consequently, by considering magnetic fields on Hubble
length scales today, our bound is a necessary condition for avoiding
backreaction. For definite primordial magnetogenesis models, the
primordial magnetic field spectrum is known such that tighter
constraints may be derived~\cite{Caprini:2001nb, Martin:2007ue,
  Subramanian:2009fu}. Up to our knowledge, the link between magnetic
field backreaction and reheating duration has only been discussed in
Ref.~\cite{Martin:2007ue} in the context of scale invariant models of
inflation breaking conformal invariance. In this work, by assuming a
magnetogenesis model, the authors use the reheating bounds coming from
CMB to get information on the allowed values of the primordial
magnetic field spectrum, its spectral index and the energy scale of
inflation. Here, we do not assume any magnetogenesis model and
consider the magnetic backreaction problem in an inverted way compared
to Ref.~\cite{Martin:2007ue}. By preventing the Hubble mode to
backreact after inflation, we extract some information on the
reheating in a model independent way. As discussed in the conclusion,
our results can also be applied to non-inflationary magnetogenesis
model provided some parameters are reinterpreted according to the
context (see Sec.~\ref{sec:conclusion}).

The paper is organized as follows. In Sec.~\ref{sec:Rradbound}, we
discuss the origin of the $\Rrad$ constraint and derive it in terms of
$\Bzero$. In Sec.~\ref{sec:cmbreh}, its implications on the energy
scale of reheating are explored by making some extra assumptions on
the equation of state parameter $\wreh$. We finally compare it with
the current WMAP7 constraints on $\Rrad$ for some specific
inflationary potentials and conclude in Sec.~\ref{sec:conclusion}.

\section{Reheating and magnetic fields backreaction}
\label{sec:Rradbound}

We assume that magnetic fields are created (for instance by quantum
vacuum fluctuations) and amplified during inflation by some mechanism
that we let unspecified. Our working hypothesis is however that
conformal invariance is restored at the end of
inflation~\cite{Turner:1987bw, Ratra:1991bn, Dolgov:1993mu,
  Dolgov:1993vg, Ratra:1994vw, Tornkvist:2000js, Gasperini:1995dh,
  Lemoine:1995vj, Bamba:2006ga, Durrer:2010mq} such that $B$ decays
subsequently as $1/a^{2}$. Therefore the strength of the field today
$\Bzero$ on large scale is given by Eq.~(\ref{eq:Bzero}). The actual
value of $\azero/\aend=1+\zend$ depends on what happens in the
universe after inflation and it is therefore related to the properties
of the reheating period.

\subsection{Reheating parameter}

Assuming instantaneous transitions between inflation,
reheating, radiation and matter era, the reheating parameter $\Rrad$
is defined by~\cite{Martin:2006rs}
\begin{equation}
\label{eq:Rraddef}
\Rrad \equiv \dfrac{a_\uend}{a_\ureh} \left( \dfrac{\rhoend}{\rhoreh}
\right)^{1/4},
\end{equation}
where ``reh'' means at the end of reheating, which is also the
beginning of the radiation era. With this definition, $\Rrad$
quantifies the deviation the reheating may have compared to a pure
radiation era ($\Rrad=1$ in that latter case, or if reheating is
instantaneous). From this definition, one can immediately evaluate the
redshift at which inflation ended
\begin{equation}
1 + \zend = (1+\zeq) \dfrac{\aeq}{\areh} \dfrac{\areh}{\aend} =
\dfrac{1}{\Rrad} \left( \dfrac{\rhoend}{\rhogamma} \right)^{1/4},
\end{equation}
where $\rhogamma$ is the energy density of radiation today\footnote{We
  have neglected a small correction eventually coming from
  non-relativistic neutrinos today and $\rhogamma=3 H_0^2 \Mpl^2
  \OmegaR$ where $\OmegaR$ the density parameter of radiation today
  ($\OmegaR \simeq 2.5 \times 10^{-5} h^{-2}$).}. Introducing the
instantaneous equation of state parameter during reheating,
$w=P/\rho$, using only energy conservation one
has~\cite{Martin:2010kz}
\begin{equation}
\label{eq:rhorehend}
\rhoreh=\rhoend \exp\left\{-3\int
_{\Nend}^{\Nreh}\left[1+w(a)\right]\ud (\ln a) \right\} = \rhoend
\ue^{-3 \Delta N (1 + \wreh)},
\end{equation}
where $\Delta N=\Nreh-\Nend \ge 0$ and $\wreh$ is the mean equation of state
parameter defined as
\begin{equation}
\wreh \equiv \dfrac{1}{\Delta N}\int_{\Nend}^{\Nreh} w(a) \ud (\ln a).
\end{equation}
Plugging Eq.~(\ref{eq:rhorehend}) into Eq.~(\ref{eq:Rraddef}) gives
back the two equivalent forms of Eq.~(\ref{eq:lnRrad}).

\subsection{Avoiding magnetic fields backreaction}

After the end of inflation the energy density of the produced magnetic
field $\rhoB$ scales as radiation. In order to avoid backreaction on
the background, we should consider two cases.

A possibility is that the reheating era has $\wreh \ge 1/3$, and the
energy density of the universe $\rho$ during reheating decays faster
than radiation such that backreaction on the length scales of interest
is avoided for
\begin{equation}
\dfrac{\rhoB(\zreh)}{\rhoreh} = \dfrac{\rhoBzero}{\rhogamma} < 1\,,
\end{equation}
which is a trivial statement. In other words, magnetic fields would
have an energy density higher than photons today. We can nevertheless
convert this bound into magnetic field values. Using Planck units
(with $\mu_0=1$), one has $1\,\G \simeq 3.3 \times 10^{-57}\Mpl^2$ and
$\rhogamma \simeq 5.7\times 10^{-125} \Mpl^4 \simeq 5.2 \times
10^{-12} \, \G^2$. Any (homogeneous) magnetic field higher than $3
\, \mu\G$ would then gravitate more than photons today, but also at
any time during the radiation era. As can be checked in
Eq.~(\ref{eq:lnRrad}), $\wreh \ge 1/3$ implies $\Rrad \ge 1$ and all
these reheating models are thus not constrained by magnetic field
backreaction.

The other possibility is $\wreh<1/3$ for which, moving back in time,
suggests that the energy density of the magnetic field $\rhoB$ may
dominate over $\rho$ during reheating. The non-trivial request that
there is no backreaction of the generated magnetic field onto the
background, namely that the energy density $\rhoB(z)$ remains smaller
than the background energy density $\rho(z)$ at any time, can be
summarized as
\begin{equation}
\label{eq:rhoBmax}
\rhoBend < \rhoend\,.
\end{equation}
A magnetic field today of value $\Bzero$ corresponds to an energy
density at the end of inflation given by
\begin{align}
\label{eq:rhoBend}
\rhoBend = \dfrac{1}{2} \dfrac{\Bzero^2}{\Rrad^4} \dfrac{\rhoend}{\rhogamma}\ .
\end{align}
From
Eqs.~(\ref{eq:rhoBmax}) and (\ref{eq:rhoBend}), we immediately get the
bound on the reheating parameter:
\begin{align}
\label{eq:constraint-reh}
\Rrad \gg \dfrac{\Bzero^{1/2} }{(2 \rhogamma)^{1/4}}\,.
\end{align}
Let us notice that Eq.~(\ref{eq:constraint-reh}) is totally
independent on the model of inflation and only requires conformal
invariance to be satisfied during the decelerating eras. Taking the
intergalactic measurements as fiducial values for the large scales
magnetic field today, Eq.~(\ref{eq:constraint-reh}) finally gives
\begin{equation}
\label{eq:lnRradmin}
\begin{aligned}
\Bzero = 5 \times 10^{-15}\,\G \quad & \Rightarrow \quad  \ln \Rrad >
-10.1\,,\\
\Bzero = 10^{-9}\,\G \quad & \Rightarrow \quad \ln \Rrad >
-4.0\,.
\end{aligned}
\end{equation}

\subsection{Reheating energy scale}

We can now plug Eq.~(\ref{eq:constraint-reh}) into
Eq.~(\ref{eq:lnRrad}) to get some information on the energy scale
and/or duration and/or the equation of state parameter of
reheating. For instance, $\rhoreh$ can be expressed in terms of
$\rhoend$, $\wreh$ and $\Rrad$ as
\begin{align}
\label{eq:rhorehRrad}
\rhoreh = \rhoend (\Rrad)^{\frac{12(1+\wreh)}{1-3\wreh}},
\end{align}
such that the above limit on $\Rrad$ provides
\begin{align}
\label{eq:limit-rhoreh}
\rhoreh > \rhoend \left[ \dfrac{\Bzero}{(2\rhogamma)^{1/2}}
\right]^{\frac{6(1+\wreh)}{1-3\wreh}}.
\end{align}
Thus, if large scale magnetic fields have an inflationary origin, this
formula tells us that once the energy scale of inflation is known,
we have a lower bound on the energy scale of reheating. From
Eq.~(\ref{eq:limit-rhoreh}), it is clear that if inflation occurs at
very low energy scale, the lower bound on $\rhoreh$ could be weaker
than what BBN already tells us ($\rhoreh > \rhonuc$). Notice that the
power goes to infinity for $\wreh \rightarrow 1/3$ such that, at fixed
$\Bzero$, the lower bound goes to zero. This is expected since, as
already mentioned, there is no non-trivial constraints from magnetic
backreaction if reheating is radiation dominated. On the other hand,
at fixed $\rhoend$, Eq.~(\ref{eq:limit-rhoreh}) shows that the lower
limit on $\rhoreh$ can be very sensitive to the value of
$\Bzero$, unless $\wreh \gtrsim -1/3$ (almost inflation).

In the next section, we compare these results to those derived using
CMB data and show that they can be tighter for some models and always
complementary.

\section{Comparison with CMB bounds}
\label{sec:cmbreh}

In inflationary cosmology, CMB anisotropies directly probe the
primordial perturbations generated during inflation. In a given
inflationary model, various CMB observables such as the spectral
index, its running or the tensor-to-scalar ratio depend on the number
of e-folds before the end of inflation at which observable wavenumbers
crossed the Hubble radius. This number depends on the
post-inflationary evolution of the universe, and thus on
reheating~\cite{Liddle:2003as, Ringeval:2005yn}. Using the WMAP seven
years data, it has been shown in Ref.~\cite{Martin:2010kz} that
reheating is actually constrained for two classes of inflationary
models, the so-called large field and small field models. For our
purpose, it is important to stress that, as discussed at length in
Ref.~\cite{Martin:2006rs}, CMB ends up being sensitive to the rescaled
parameter
\begin{equation}
\label{eq:Rdef}
R \equiv \Rrad \dfrac{\rhoend^{1/4}}{\Mpl}\,,
\end{equation}
which involves an extra factor $\rhoend^{1/4}$ compared to the reheating
parameter.

\subsection{Large field models}

\subsubsection{Generic reheating}

In large field models, the inflationary potential has the following
form
\begin{align}
\label{eq:lfpot}
V(\phi) = M^{4} \left( \frac{\phi}{\Mpl} \right)^{p},
\end{align}
where $M$ is the energy scale which fixes the amplitude of the CMB
anisotropies and $p$ is a free index. For large field models, both $R$
and $\rhoend$ have been shown to be constrained, and after
marginalization over $0.2<p<5$ and over the standard cosmological
parameters, Ref.~\cite{Martin:2010kz} reports the two-sigma confidence
intervals from WMAP data:
\begin{equation}
\label{eq:lfcmb}
\ln R > -28.9, \qquad 4
\times 10^{15}\, \GeV < \rhoend^{1/4} < 1.2 \times 10^{16} \, \GeV.
\end{equation}
From this equation, taking the central value $\rhoend^{1/4} \simeq 8
\times 10^{15} \,\GeV$ we get from Eq.~(\ref{eq:Rdef}) that $\ln \Rrad
> -23$. This has to be compared to the magnetic field limit of
Eq.~(\ref{eq:lnRradmin}). As a result, reheating in large field models
is currently more constrained by the magnetic fields bound.

\subsubsection{Assuming an equation of state}

Although Eq.~(\ref{eq:lfcmb}) does not make any assumptions on the
reheating, large field models are expected to end with parametric
oscillations around the minimum of the potential. In that case, the
equation of state parameter during reheating reads
$\wreh=(p-2)/(p+2)$~\cite{Turner:1983he, Kofman:1997yn, Liddle:2003as,
  Martin:2003bt}. Let us derive the bound on the energy scale of
reheating given by inflationary magnetic fields for different choices
of $\wreh$.

One can argue that, by Taylor expanding the potential, and for small
enough field values, any potential should behave as $\phi^2$. In that
case, parametric oscillations proceed with $\wreh \simeq 0$ and the
reheating era expands as a matter era~\cite{Turner:1983he}. Using CMB
to get $\rhoend$ given above, assuming $\wreh=0$, the magnetic bound
of Eq.~(\ref{eq:limit-rhoreh}) gives
\begin{equation}
\label{eq:rhorehmat}
\begin{aligned}
\Bzero =5\times 10^{-15}\,\G \quad & \Rightarrow \quad \rhoreh^{1/4} >
490 \,\GeV, \\
\Bzero = 10^{-14}\,\G \quad & \Rightarrow \quad \rhoreh^{1/4} >
1.4 \times 10^{3}\,\GeV, \\
\Bzero = 10^{-9}\,\G \quad & \Rightarrow \quad \rhoreh^{1/4} >
4.3 \times 10^{10} \,\GeV.
\end{aligned}
\end{equation}

Another possible choice is to assume that symmetries ensure that the
potential (\ref{eq:lfpot}) is valid at all field values, then one has
$\wreh=(p-2)/(p+2)$ for all $p$. Taking a negative $\wreh$ actually
yields stronger bounds. For instance, with the quite extreme value
$\wreh = -0.3$, we find
\begin{equation}
\label{eq:rhorehinf}
\begin{aligned}
\Bzero=5\times 10^{-15}\,\G \quad &\Rightarrow \quad \rhoreh^{1/4} > 1.1 \times
10^{11}\,\GeV, \\
\Bzero = 10^{-14}\,\G \quad & \Rightarrow
\quad \rhoreh^{1/4} > 1.6 \times 10^{11} \,\GeV, \\
\Bzero = 10^{-9}\,\G \quad & \Rightarrow \quad \rhoreh^{1/4} > 9.1
\times 10^{13}\,\GeV.
\end{aligned}
\end{equation}
As already noticed, the sensitivity to the $\Bzero$ values goes down
when $\wreh$ approaches acceleration ($\wreh \rightarrow -1/3$).

The numbers of Eq.~(\ref{eq:rhorehinf}) cannot be straightforwardly
compared to those coming from CMB alone as Ref.~\cite{Martin:2010kz}
gives only marginalized results over all values of $p$. In order to be
consistent, we have reproduced the same CMB analysis as in
Ref.~\cite{Martin:2010kz} fixing the power $p$ of the potential at
either $p=2$ or $p \gtrsim 1$ (corresponding to $\wreh=0$ or $\wreh
\gtrsim -0.3$, respectively). For the sake of clarity, we give only
the two numbers we are interested in\footnote{The CMB bounds on
  $\rhoend$ remain typically the same.}. For $p=2$, and thus
$\wreh=0$, we find the two-sigma CMB lower bound $\rhoreh^{1/4} >
70\,\GeV$, which ends up being less stringent than all magnetic bounds
of Eq.~(\ref{eq:rhorehmat}). Doing the same for $p \gtrsim 1$ and
therefore $\wreh \gtrsim -0.3$, CMB gives $\rhoend^{1/4} > 10^{6}
\,\GeV$ which is again less stringent than Eq.~(\ref{eq:rhorehinf}).

\subsection{Small field models}

Small field models have a potential given by
\begin{equation}
\label{eq:sfpot}
V(\phi) = M^4 \left[1 - \left(\dfrac{\phi}{\mu} \right)^p \right]\,,
\end{equation}
and the CMB data analysis with WMAP7 data gives the lower two-sigma
limit~\cite{Martin:2010kz}
\begin{equation}
\label{eq:sfcmb}
\ln R > -23.1\,.
\end{equation}
Interestingly, for a generic reheating, there is no CMB bound on
$\rhoend$ for small field models such that $\Rrad$ alone remains
unconstrained. As a consequence, Eq.~(\ref{eq:constraint-reh}) is
complementary since it gives some information on $\Rrad$. From
Eq.~(\ref{eq:Rdef}), one has $\ln R=\ln \Rrad + (1/4)\ln
(\rhoend/\Mpl^4)$ and the magnetic bound of Eq.~(\ref{eq:lnRradmin})
would be stronger than CMB only if $\rhoend^{1/4} > 6 \times
10^{12}\,\GeV$ (for fiducial $\Bzero=5\times 10^{-15}\,\G$). If the
energy scale of inflation is lower than this value, CMB remains more
restrictive than avoiding magnetic backreaction, as expected from the
previous discussions.

Using the highest possible value of $\Bzero = 10^{-9}\,\G$, the
magnetic bound becomes stronger than CMB limits only if $\rhoend^{1/4}
> 1.3 \times 10^{10}\, \GeV$. Conversely, there is no magnetic field
backreaction problem on Hubble scale today in all small field models
fitting CMB and having an energy scale lower than this value. In that
case, Eq.~(\ref{eq:constraint-reh}) is automatically verified.

\section{Conclusion}
\label{sec:conclusion}

We have shown in this paper that avoiding magnetic field backreaction
yields some non-trivial constraints on the reheating epoch. Our
results have been derived under the assumption that conformal
invariance is restored after inflation and can be summarized in terms
of the reheating parameter $\Rrad$ as $\ln \Rrad > -10.1$ for $\Bzero
= 5\times 10^{-15}\,\G$ [see Eq.~(\ref{eq:constraint-reh})].

Compared and combined with CMB data, we have shown that these
conditions translate into a lower bound for the large field reheating
energy scale when one assumes $\wreh$ to be known. For any large field
model having a matter dominated reheating, we find the lower limit for
$\rhoreh^{1/4}$ ranging from $490\,\GeV$ to $4.3 \times 10^{10}\,\GeV$
for magnetic field values varying from $\Bzero = 5\times 10^{-15}\,
\G$ to $10^{-9}\,\G$. For the small field models, we find that
avoiding magnetic backreaction is more constraining than CMB only if
the energy scale of inflation remains large enough, typically higher
than $10^{12} \,\GeV$. In fact, as suggested by Eq.~(\ref{eq:Bzero}),
magnetic field backreaction is all the more important for $\zend$
large. For this reason, the magnetic bounds are expected to be
important on all high energy inflationary models while being easily
satisfied for those at low energy. Let us stress again that,
although model independent, our bound is a necessary condition in the
sense that it must be always satisfied. According to the shape of the
primordial magnetic field spectrum, backreaction may eventually be
stronger on smaller length scales. In this case, if the spectrum is
known, Eq.~(\ref{eq:rhoBmax}) still applies provided one uses the
magnetic field value at the scale the spectrum is maximal.

Finally, the above analysis can be straightforwardly generalized to
any non-conventional post-inflationary thermal history. As discussed
in Ref.~\cite{Martin:2010hh}, if the universe evolution incorporates a
new $X$-era, in addition to the reheating, one can define a new
parameter $\Rx$ exactly as in Eq.~(\ref{eq:Rraddef}). All of our
previous results would still apply to the combination $\Rrad \Rx$
instead of $\Rrad$. In fact, Eq.~(\ref{eq:constraint-reh}) is equally
applicable to any model of primordial magnetogenesis having a
decelerating era preserving conformal invariance and occurring before
radiation domination. In that case, $\rhoend$ has to be understood as
the energy density of the universe when this era starts and $\rhoreh$
when it ends.

\acknowledgments

It is a pleasure to thank Jerome Martin for suggestions and
enlightening comments on the manuscript. This work is partially
supported by the Wallonia-Brussels Federation grant ARC 11/15-040. VD
is supported by ESA under the Belgian Federal PRODEX program
$\mathrm{N}^\circ 4000103071$.

\bibliographystyle{JHEP}
\bibliography{magneticfields,reheating}

\end{document}